\documentclass[12pt,preprint]{aastex}

\slugcomment{accepted by AJ}

\shorttitle{Thermal Modeling of Near-Earth Asteroids} 
\shortauthors{Mommert, Jedicke, Trilling}

\begin{document}


\title{An Investigation of the Ranges of Validity of Asteroid Thermal
  Models for Near-Earth Asteroid Observations}


\author{M. Mommert\altaffilmark{1}}

\author{R. Jedicke\altaffilmark{2}}

\author{D.\ E.\ Trilling\altaffilmark{1}}


\altaffiltext{1}{Department of Physics and Astronomy, Northern Arizona
  University, Flagstaff, AZ 86011, USA} 
\altaffiltext{2}{Institute for Astronomy, University of Hawai\'{}i at
  Manoa, Honolulu, HI 96822, USA}


\begin{abstract}
  The majority of known asteroid diameters are derived from thermal
  infrared observations. Diameters are derived using asteroid thermal
  models that approximate their surface temperature distributions and
  compare the measured thermal infrared flux with model-dependent
  predictions. The most commonly used thermal model is the Near-Earth
  Asteroid Thermal Model \citep[NEATM,][]{Harris1998}, which is
  usually perceived superior to other models like the Fast-Rotating
  Model \citep[FRM,][]{Lebofsky1989}. We investigate the applicability
  of the NEATM and the FRM to thermal-infrared observations of
  Near-Earth Objects using synthetic asteroids with properties based
  on the real Near-Earth Asteroid (NEA) population. We find the NEATM
  to provide more accurate diameters and albedos than the FRM in most
  cases, with a few exceptions. The modeling results are barely
  affected by the physical properties of the objects, but we find a
  large impact of the solar phase angle on the modeling results. We
  conclude that the NEATM provides statistically more robust diameter
  estimates for NEAs observed at solar phase angles less than
  ${\sim}$65\degr, while the FRM provides more robust diameter
  estimates for solar phase angles greater than ${\sim}$65\degr. We
  estimate that ${<}5$\% of all NEA diameters and albedos derived up
  to date are affected by systematic effects that are of the same
  order of magnitude as the typical thermal model uncertainties. We
  provide statistical correction functions for diameters and albedos
  derived using the NEATM and FRM as a function of solar phase angle.
\end{abstract}





\section{Introduction}

Within the last decade, thermal-infrared observations and thermal modeling
of asteroids have increased the number of asteroids with measured
diameters and albedos by orders of magnitude. This increase can be
mostly attributed to a new generation of space-based infrared
telescopes that have been utilized in a number of asteroid discovery
and characterization surveys: the {\it Spitzer Space Telescope}
\citep{Werner2004} has been used in the ExploreNEOs
\citep{Trilling2010}, NEOSurvey \citep{Trilling2016}, and ongoing NEOLegacy
characterization surveys, among many other asteroid-related
observations; {\it Akari} \citep{Usui2011} all-sky observations have
been used to establish the ``Asteroid catalog using Akari''; and the
{\it Wide-field Infrared Survey Explorer} \citep[WISE,][]{Wright2010}
has been used in the NEOWISE program \citep{Mainzer2011b} in an
all-sky discovery and characterization survey. More than 100,000 main
belt asteroids and about 2,500 near-Earth asteroids (NEAs) with
perihelion distances $q \leq 1.3$~au have measured diameters and
albedos using this method \citep[also see][for a
discussion]{Mainzer2015}.

\begin{figure}
\epsscale{0.6}
\plotone{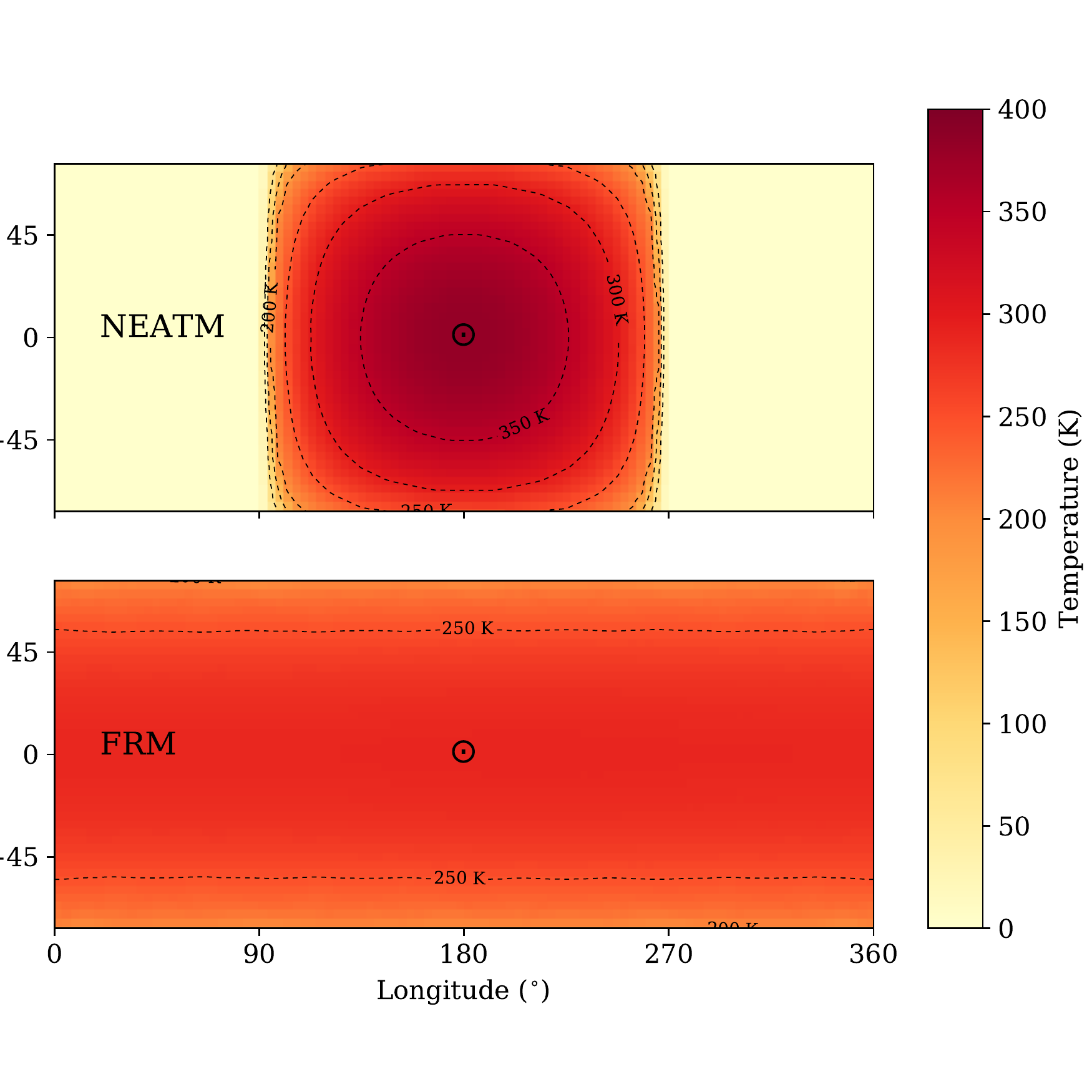}
\plotone{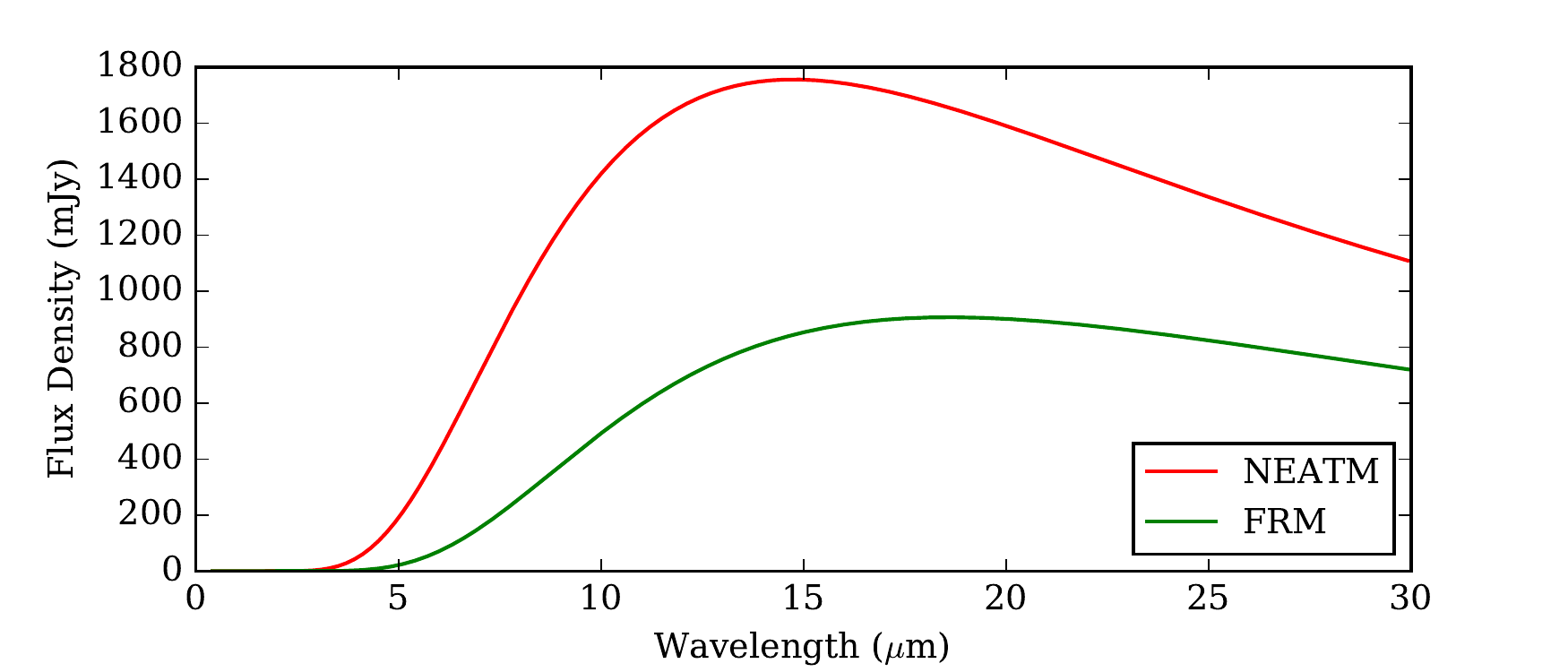}
\caption{Representation of the surface temperature distributions
  assumed by the NEATM (top, using beaming parameter $\eta=1.0$) and
  FRM (center) for the same asteroid 1~au from the Sun. The subsolar
  point (Sun symbol) coincides with the sub-observer point in both
  plots. Contour lines refer to areas with similar temperatures. Note
  that the maximum temperature of the FRM is significantly lower than
  that of the NEATM. The bottom panel shows the spectral energy
  distribution (SED) caused by thermal emission for both models. The
  NEATM's higher temperatures cause a shift of its SED peak towards
  shorter wavelengths as compared to the
  FRM.\label{fig:neatm_frm_temps}}
\end{figure}

The majority of asteroid thermal data is interpreted using radiometric
or thermal models.  Asteroid thermal models combine thermal-infrared
and optical brightness measurements in a simplified representation of
the asteroid's surface temperature distribution that allows for a
derivation of the asteroid's diameter and albedo. The most widely used
thermal model is the Near-Earth Asteroid Thermal Model
\citep[NEATM,][]{Harris1998}. The NEATM is based on the earlier
Standard Thermal Model \citep[STM, e.g.,][]{Morrison1979}, both of
which assume a spherical shape of the model asteroid, a spin axis
orientation perpendicular to the orbital plane, a very slowly rotating
asteroid, and a Lambertian surface with a negligible thermal
inertia. The resulting temperature distribution (see Figure
\ref{fig:neatm_frm_temps}, top) consists of concentric small circles
of the same temperature centered on the sub-solar point; both models
assume zero night-side thermal emission. The NEATM fully accounts for
the solar phase angle of the observations, whereas the STM assumes the
target to be observed at zero phase angle (STM and NEATM are identical
in the zero phase angle case); an empirical phase correction for
thermal-infrared fluxes was derived for the STM by
\citet{Lebofsky1989} to compensate for this simplification. The NEATM
also allows for a modulation of the surface temperature distribution
with a variable beaming parameter $\eta$; a fixed value of $\eta$ was
introduced in the ``refined STM'' from observations by
\citet{Lebofsky1986}. The beaming parameter accounts for thermal
effects caused by a non-zero thermal inertia of the surface material,
surface roughness, and the spin state of the real asteroid in a
zero-th order approximation. The NEATM has been found to provide
robust diameter and albedo estimates \citep{Delbo2004, Wright2007,
  Harris2011, Mainzer2011a} and is usually preferred over the STM due
to its flexibility. \citet{Harris2011} find NEATM diameters and
albedos to be accurate within 20\% and 50\%, respectively.

A different thermal model that deviates from the definition of the
NEATM in assuming a rapid rotation of the model asteroid and a very
high thermal inertia of the surface material is the Fast-Rotating
Model \citep[FRM,][]{Lebofsky1989}. As a result of the body's fast
rotation, insolation is evenly distributed over the sphere and the
resulting surface temperature distribution is iso-latitudinal; the
equator is hottest and the temperature decreases towards the poles of
the model asteroid (see Figure \ref{fig:neatm_frm_temps}, center).

STM/NEATM and FRM represent two extreme cases: STM/NEATM assume no (or
very slow) rotation and zero (or a very small) thermal inertia,
whereas the FRM assumes a fast rotation and high thermal inertia. For
the majority of asteroids, the NEATM performs significantly better
than the FRM \citep{Delbo2004}, which is due to its increased
flexibility through the variable beaming parameter $\eta$, as well as
the extreme assumptions characterizing the FRM. Generally, high
thermal inertias and/or rotation rates are necessary to generate a
surface temperature distribution similar to that of the FRM. However,
\citet{Delbo2004} found that the FRM produces more accurate diameter
results than the NEATM for large phase angles.

Thermal modeling of NEAs technically does not differ from the modeling
of any other type of atmosphereless body. However, there are some
aspects of the properties of NEAs that should be taken into account to
properly interpret their thermal modeling results. NEAs are closer to
the Earth than other Solar System bodies. Hence, they can be observed
at larger solar phase angles (the angle between the Sun and the
observer as measured from the asteroid.)  While the NEATM
geometrically accounts for the solar phase angle, the zero night-side
temperature assumed by the NEATM compromises diameter and albedo
measurements with increasing solar phase angle
\citep[e.g.,][]{Delbo2004, Wright2007, Wolters2009}. In some extreme
cases -- for instance fast rotation or high thermal inertia -- the
observed temperature distribution might be better represented by the
FRM than the NEATM. Such scenarios are relevant since NEAs tend to
have shorter rotational periods than larger and more distant asteroids
\citep[see rotational periods compiled by][]{Warner2009}. The current
record-holder for the fastest-spinning asteroid, 2014~RC, completes
one rotation in only 16~s \citep{Moskovitz2015}. As a result of their
fast rotation, the FRM might provide more realistic diameters and
albedos than the NEATM for these objects. This is especially relevant
for the smallest NEAs to be discovered in the future, among which will be 
temporarily captured Earth satellites or minimoons \citep{Granvik2012,
  Bolin2014} that come extremely close to Earth.



Previous studies investigating the accuracy of the NEATM applied to
observations of NEAs have been conducted. \citet{Delbo2004} and
\citet{Wright2007} utilized a comparison of the NEATM with asteroid
thermophysical models to investigate the effects of solar phase angle,
thermal inertia, and surface roughness on the NEATM accuracy. In this
work, we elaborate on their approach and perform a more detailed and
quantitative analysis of these effects on the modeled diameters and
albedos as a function of asteroid physical properties and observation
geometries. We consider conservative ranges for the physical
properties based on our current picture of the NEA population and
identify asteroid configurations under which the FRM is superior to
the NEATM and provide a statistical analysis of the systematic effects
on diameters and albedos derived with the NEATM and the FRM. The goal
of this work is to provide a guideline for when thermal modeling
results are reliable -- or not -- and how they should be interpreted.

The results of this work have important implications for the {\em
  a-priori} use of the NEATM in large-scale characterization programs
and the thermal modeling of the smallest and closest NEAs.

\section{Methodology}

We explore the applicability of the NEATM and the FRM by simulating
thermal-infrared flux densities for 1 million synthetic NEAs using an
asteroid thermophysical model (TPM, Section \ref{lbl:fluxes}) in a
Monte Carlo approach. The physical properties and orbital geometries
of these synthetic NEAs follow distributions that are typical for NEAs
according to our best knowledge (Section \ref{lbl:space}). The
simulated flux densities are modeled with NEATM and FRM
implementations to derive diameter and albedo estimates (Section
\ref{lbl:modeling}), which are subsequently compared against each
other and the simulation input parameters (Section
\ref{lbl:analysis}).

\subsection{Simulated Flux Densities}
\label{lbl:fluxes}

We simulate thermal flux densities using a detailed asteroid
thermophysical model. Thermophysical models (TPMs) allow for
simulating the asteroid's surface temperature distribution by solving
the heat transfer equation numerically for a large number of plane
surface facets. The TPM we use is mostly identical to, and has been
tested extensively against, the one discussed by \citet{Mueller2007}
and has been used in other works, including \citet{Mommert2014a} and
\citet{Mommert2014b}. Our TPM implementation accounts for the spin
axis orientation (longitude and latitude in ecliptic coordinates),
rotational period, thermal inertia, and surface roughness in addition
to observing geometry. TPMs can provide more accurate diameter and
albedo estimates than simple thermal models, but they also require
additional constraints on the aforementioned parameters. In this work,
we use the TPM to simulate flux densities for synthetic asteroids.

The TPM solves the heat transfer equation numerically for each surface
facet until the sub-surface temperature at a certain depth varies less
than a certain fractional threshold, $\varepsilon_{\mathrm{temp}}$
(usually $\varepsilon_{\mathrm{temp}} = 10^{-4}$), between two full
rotations. For the sake of simplicity, we use a spherical shape for
our model asteroids, each consisting of 496 triangular surface
facets. Surface roughness is taken into account and modeled as
emission from spherical craters using the approach taken by
\citet{Mueller2007}; different models (no roughness, low roughness,
default roughness, and high roughness) as defined in
\citet{Mueller2004} are utilized here.  In order to verify the
conservation of energy in the model, the amount of incoming and
outgoing energy are compared against each other. If there is an energy
discrepancy of 1\% or greater, $\varepsilon_{\mathrm{temp}}$ is
reduced by an order of magnitude. This adjustment improves the
conservation of energy and leads to more accurate flux densities --
but also requires more computation time.

The wavelengths at which the flux densities are simulated sample the
peaks of the asteroids' spectral energy distributions well:
5--20~$\mu$m at intervals of 5~$\mu$m. Note that reflected solar light
is not considered as part of this work.  Flux density uncertainties,
which are necessary for the proper interpretation of the flux
densities as part of the $\chi^2$ minimization in the thermal models
(see Section \ref{lbl:modeling}), are assumed to follow a Poisson
distribution and hence are the square-root of the derived flux at each
wavelength (in order to mimic the noise properties of real
observations). 

\subsection{Parameter Space}
\label{lbl:space}

We create a synthetic representation of the NEA population consisting
of 1 million objects sampling the parameters of our TPM (see Section
\ref{lbl:fluxes}). The ranges we consider for the physical properties
are typical for NEAs and we weight them according to their measured or
modeled distributions. Orbital parameters are chosen from the
population of known NEAs, as detailed below.

In order to simulate flux densities, we have to assume diameters and
geometric albedos for our synthetic NEAs. Geometric albedos ($p_V$)
are drawn from the albedo distribution by \citet{Wright2016} measured
for the NEA population during the cryogenic part of the WISE
mission. Since the surface temperature distribution of an asteroid is
independent of its size, we assume a constant absolute magnitude
$H_V = 18$~mag ($G = 0.15$) for all objects; the target diameter
follows from $H_V$ and the respective geometric albedo value.

We draw ecliptic spin axis latitudes from a second-order polynomial
distribution fitted to the obliquity distribution derived by
\citet{Farnocchia2013} (their Figure 6); this distribution is in
agreement with later findings \citep{Tardioli2017}. Spin axis
longitudes are drawn from a uniform distribution spanning 360\degr\ as
there is no preference for spin axis longitudes. Other parameters are
also drawn from uniform distributions due to a lack of additional
information on the underlying distributions. We sample thermal
inertias in the range $\Gamma \in [20, 2000]$~SI units (1~SI Unit =
1~J~m$^{-2}$~s$^{-0.5}$~K$^{-1}$), which is motivated by
\citet{Mommert2014a} and the fact that little is known about the
un-biased distribution of NEA thermal inertias.  Rotational periods
are uniformly sampled from the interval
$P \in [10~\mathrm{sec}, 100~\mathrm{hr}]$; this range is consistent
with the range of reliably measured rotational periods of NEAs
compiled by \citet{Warner2009}. We note that the distribution of
rotational periods of NEAs and other asteroids is strongly biased
\citep[e.g.,][]{Masiero2009}. Hence, we deliberately use a uniform
distribution of rotational periods in order to probe the impact of
rotational periods over a wide range. Finally, we uniformly sample the
discrete roughness models [low, default, and high roughness]
\citep[see][for definitions of these models]{Mueller2007}; we consider
the absence of surface roughness unrealistic for this study. We refer
to Section \ref{lbl:discussion} for a discussion of the range
definitions and their implications.

In order to properly sample observation geometry, we use heliocentric
and geocentric coordinates of 5,000 randomly selected real NEAs for a
random epoch. All selected NEAs
have solar elongation angles $60 \leq \varepsilon \leq 180\degr$,
restricting the sample to objects that are observable from the Earth
and from space-based platforms with state-of-the-art thermal infrared
telescopes. Hence, our set of orbital geometry parameters provides a
snapshot of the NEA population that is accessible for observations. We
verified that this subset of the NEA population represents the overall
population well in heliocentric and geocentric distances, solar phase
angle, and as a function of time (using different epochs). This
verification was performed using a two-sided Kolmogorov-Smirnov test
that compares the full ensemble of observable NEAs at different epochs
with our random sample; in no case was there a significant probability
that our sample was drawn from a different population.

For each of the 5,000 NEA geometries, we create 200 randomized sets of
physical properties, resulting in a total of 1 million synthetic NEAs
considered in this study. We do not consider noise contributions to
the sampled parameters in order to be able to better work out the
effects at work.


\subsection{Thermal Modeling}
\label{lbl:modeling}

The simulated flux densities are modeled using well-tested versions of
the NEATM and the FRM, following the implementation by
\citet{Delbo2002}. For each synthetic NEA, we use all four generated
thermal flux densities and fit them simultaneously taking into account
the observing geometry and flux density uncertainties. The best fit is
derived using a $\chi^2$ minimization of the residuals between
``measured'' (generated by the TPM) and thermal model flux densities,
weighted by the ``measured'' flux density uncertainties. The use of
realistic (Poissonian) ``measured'' flux density uncertainties is
necessary to weight the ``measured'' flux densities at the different
wavelengths correctly. Improper uncertainties will lead to distorted
best-fit spectral energy distributions and hence systematic effects on
the derived diameters and albedos. Using this approach, we fit
diameters and beaming parameters (NEATM only) for each object --
albedos follow through the combination with the assumed absolute
magnitude of the object. We use the same Solar System absolute
magnitude as a measure for optical brightness for all objects, which
does not affect the conclusion of this work. We refer to
\citet{Mommert2013} for a detailed discussion of the model
implementations.

\subsection{Analysis}
\label{lbl:analysis}


For each synthetic object we can identify whether the NEATM or the FRM
provides the more accurate diameter or albedo result by comparing our
solutions to the input parameters utilized in the derivation of the
flux densities using the TPM (see Section \ref{lbl:fluxes}). For
instance, the diameter derived with the NEATM is considered more
accurate if $|d_{\mathrm{NEATM}} - d| < |d_{\mathrm{FRM}}- d|$, where
$d$ is the input diameter, $d_{\mathrm{NEATM}}$ is the diameter
derived by the NEATM, and $d_{\mathrm{FRM}}$ is the diameter derived
by the FRM.  In order to characterize the impact of one specific
physical or orbital parameter on diameter and albedo estimates, we
marginalize over the other parameters and derive distributions where
the NEATM (or the FRM) is more accurate. The {\bf success rate } is
defined as the fraction of cases where the NEATM (or FRM) provides a
better fit over a marginalized distribution. The marginalized
distribution is derived from the complete synthetic object
distribution by summation over the success rates for all but one
parameter. Hence, the success rate distribution for this one parameter
is fully retained, whereas information from the other parameters has
been marginalized out.  For instance, the NEATM diameter success rate
with respect to thermal inertia describes the fraction of cases in
which the NEATM derives a more accurate diameter than the FRM as a
function of thermal inertia; in this case, the success rates for
  all parameters except that of the thermal inertia are summed over,
  retaining the success rate distribution with respect to this
  parameter. This fraction can be interpreted as the probability of
the NEATM to be superior to the FRM as a function of thermal inertia
for a random asteroid. Success rates for the diameter and the albedo
should be identical; we provide both quantities for the sake of
completeness and discuss exceptions from this rule.

In order to quantify the agreement of thermal model results with the
actual synthetic asteroid diameters and albedos, we introduce a second
diagnostic quantity. For each synthetic asteroid, we calculate the
quantity $(d_{\mathrm{NEATM}}-d)/d$, which is the fractional error of
the NEATM-derived diameter ($d_{\mathrm{NEATM}}$) from the actual
input diameter ($d$); errors for the FRM and for the geometric albedo
can be defined correspondingly. By marginalizing over one parameter we
can define the {\bf median fractional error} as a function
of this parameter. The marginalization is done as described
  above. However, in this case the median is derived instead of the
  sum of the other parameters. The marginalization enables the
identification of systematic effects that occur as a function of
individual parameters. The variation in the fractional errors is
captured by the 1$\sigma$ and 3$\sigma$ envelopes that include 68.3\%
and 99.7\% of all synthetic objects around the median in each bin,
respectively.

\section{Results}
\label{lbl:results}

The following sections investigate the accuracy of the NEATM and the
FRM as a function of different synthetic object parameters using the
tools defined in Section \ref{lbl:analysis}.

\subsection{Physical Properties}

\begin{figure}
\epsscale{0.6}
\plotone{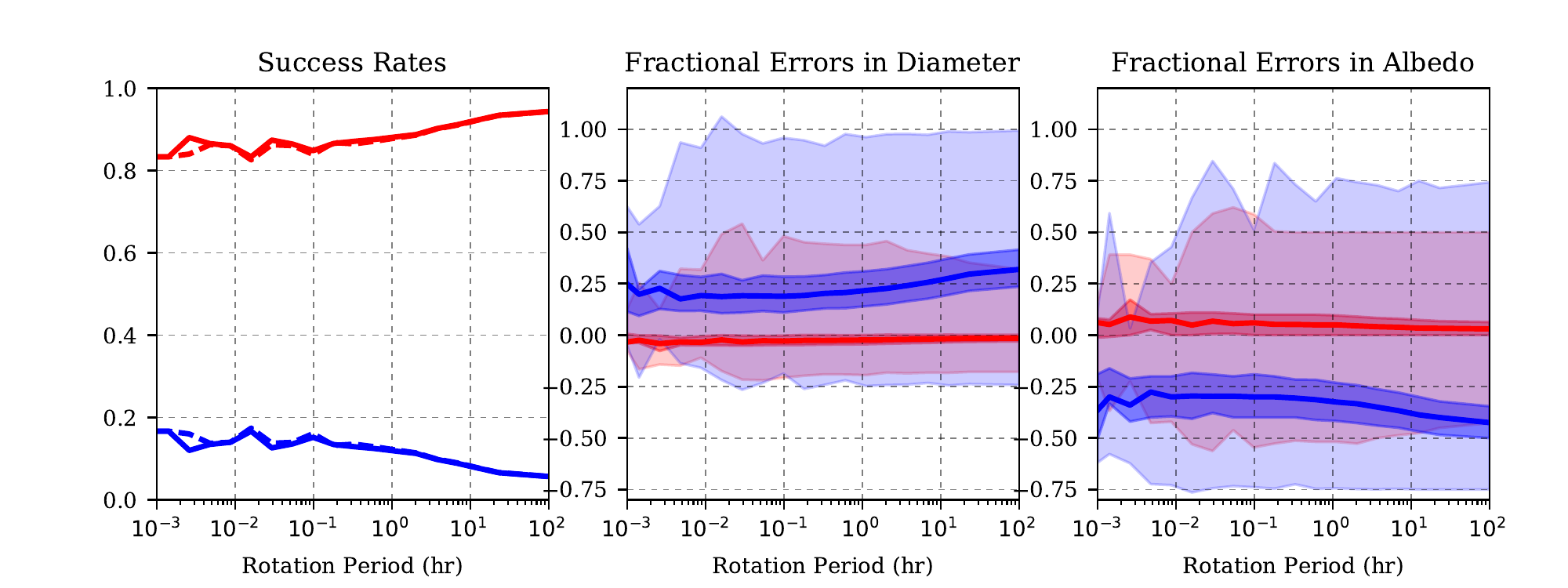}
\plotone{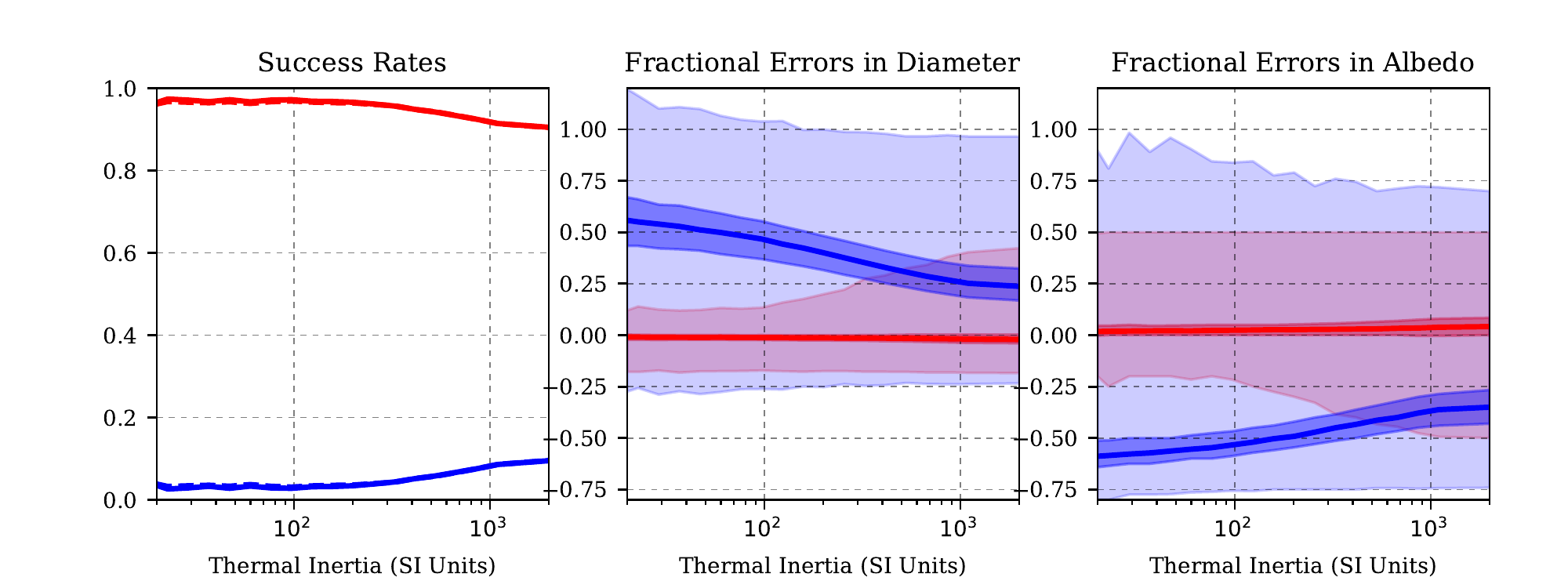}
\plotone{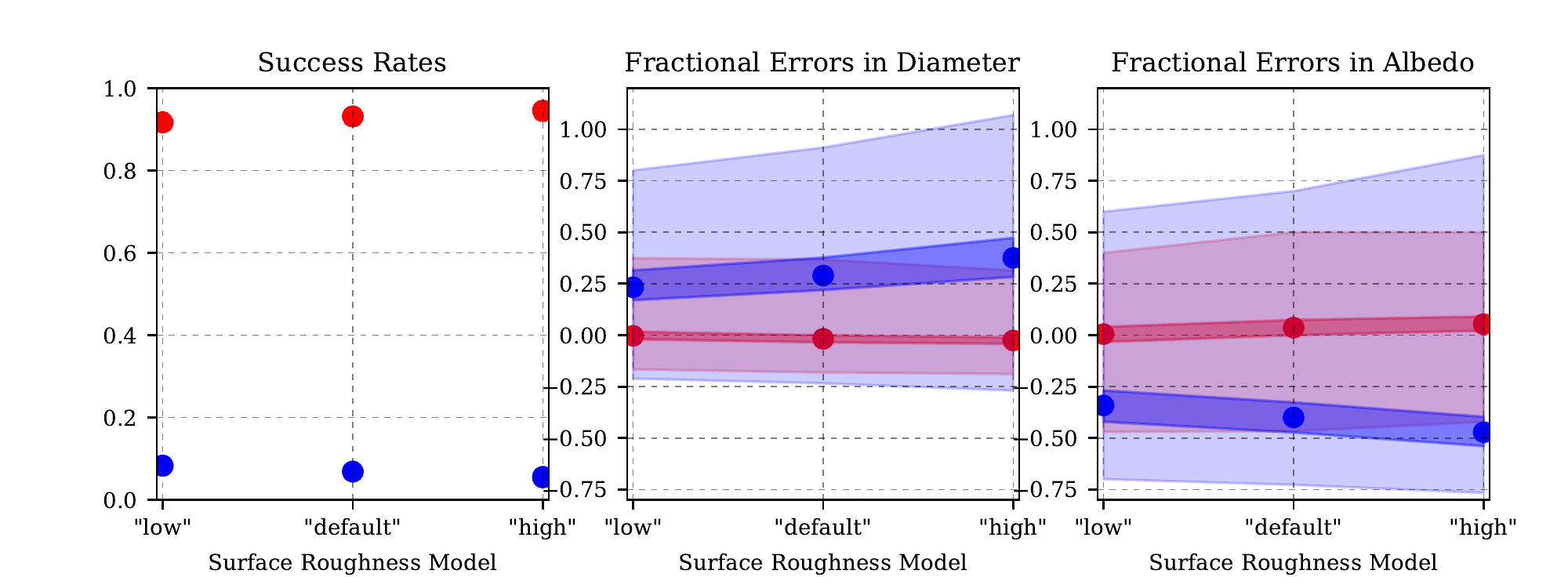}
\plotone{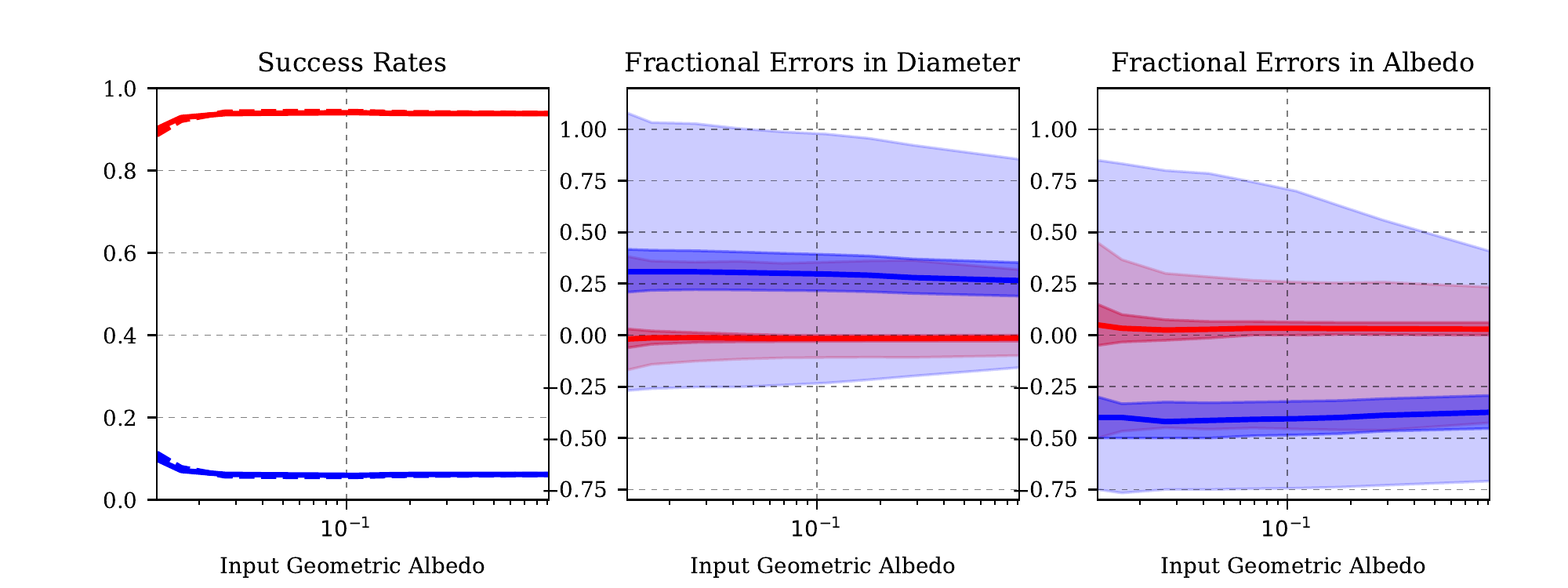}
\caption{Success rates and fractional errors based on physical
  properties: rotation period (top row), thermal inertia (second from
  top), surface roughness (second from bottom), and geometric albedo
  (bottom row). In all plots, NEATM is indicated with red and FRM is
  indicated with blue. The left columns show NEATM and FRM success
  rates for diameter (solid line) and albedo (dashed line); the center
  column shows diameter median fractional errors (solid line) as well
  as its 1$\sigma$ (darker shading) and 3$\sigma$ (lighter shading)
  envelopes; the right column shows albedo fractional errors.  For all
  properties considered here, the NEATM is more likely to provide
  accurate diameters and albedos than the FRM. NEATM median fractional
  errors are negligible, while FRM median fractional errors are
  significant.\label{fig:physicalproperties}}
\end{figure}

Figure \ref{fig:physicalproperties} shows success rates and fractional
errors for rotational period, thermal inertia, surface roughness,
and geometric albedo. NEATM success rates are greater than 80\% and the
NEATM median fractional errors are zero within the 1$\sigma$
intervals for all parameters, whereas the FRM fractional errors are
systematically and significantly offset. Conditions that resemble the
assumptions of the FRM (high thermal inertia, short rotation periods)
lead to slightly higher success rates and lower fractional errors
for the FRM and vice-versa for the NEATM. However, this effect is not
significant. Increasing degrees of surface roughness lead to
increasing fractional errors for both the NEATM and the FRM, but
with different signs. This is expected since both the NEATM and the
FRM assume Lambertian surfaces that are perfectly smooth. Finally,
diameters and geometric albedos derived with both models are
unaffected by the input geometric albedos of the synthetic objects.

The physical properties have only minor effects on the applicability
of the NEATM, which is superior to the FRM in the vast majority of
cases.

\subsection{Spin Axis Orientation}

\begin{figure}
\epsscale{1}
\plotone{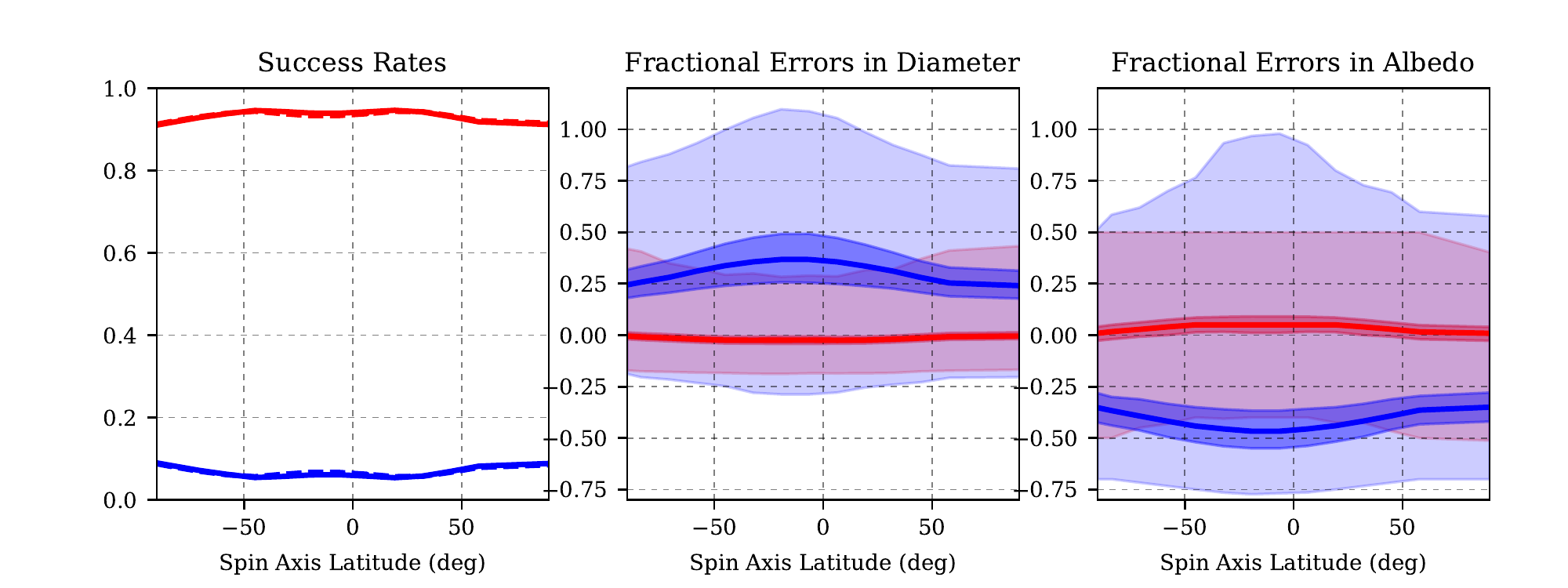}
\plotone{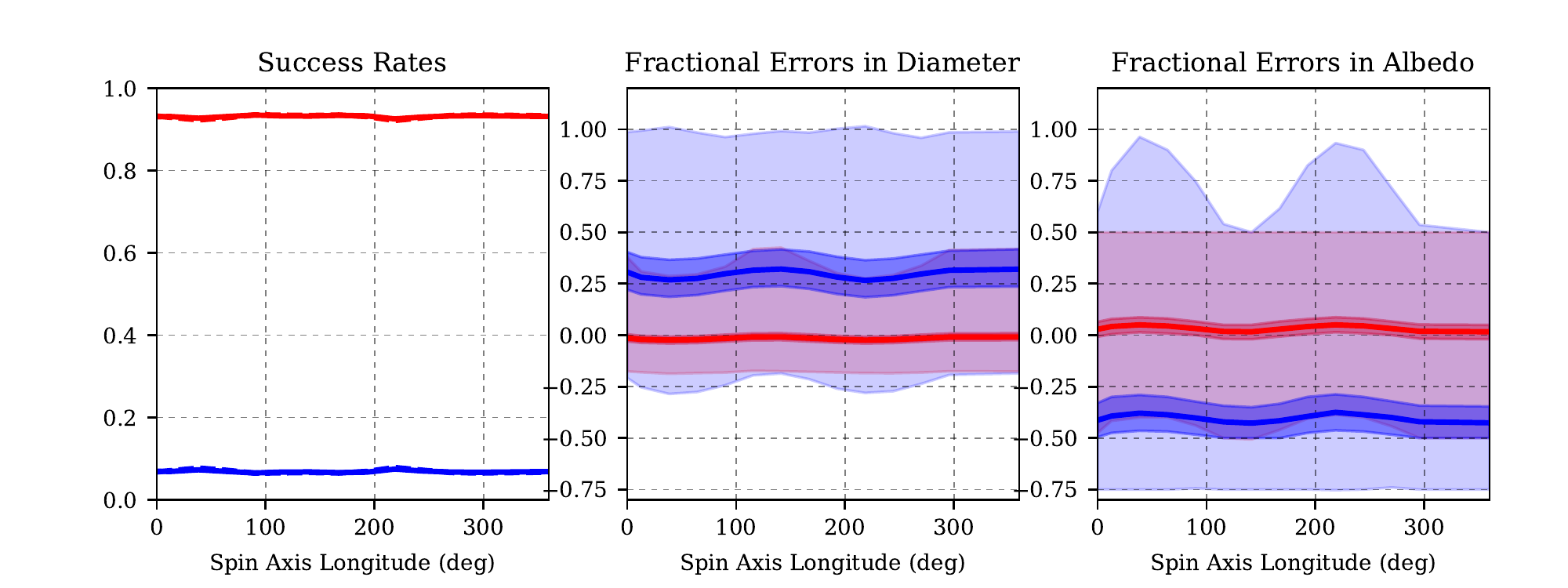}
\caption{Success rates and fractional errors based on spin axis
  latitudes and longitudes. Refer to Figure
  \ref{fig:physicalproperties} for definitions. Based on the success
  rates and fractional errors, the NEATM is more likely to provide
  accurate diameters and albedos than the
  FRM. \label{fig:spinaxis}}
\end{figure}

The impact of the spin axis orientation on thermal modeling is shown
in Figure \ref{fig:spinaxis}. The NEATM success rates are greater 90\%
and NEATM fractional errors are negligible throughout the entire
range in both parameters. While the fractional errors for the spin
axis latitude show a dip/bump, the fractional errors for the spin
axis longitude show a wavy pattern. The former effect, a dip in the
fractional errors of the diameters (albedos) derived with the NEATM
(FRM) and a bump in the fractional errors of the diameters
(albedos) derived with the FRM (NEATM) at spin axis latitudes close to
zero, is a result of the different surface temperature distributions
assumed by the two thermal models that are more or less compatible
with equator-on views. Similarly, the NEATM has higher success rates
for spin axis latitudes around zero and the FRM has higher success
rates for extreme spin axis latitudes (-90\degr and +90\degr). At spin
axis latitudes close to zero (and a random spin axis longitude), it is
somewhat likely to have insolation and hence higher temperatures at
the poles of the object, mimicking the NEATM surface temperature
distribution and preferring the NEATM. In this case, the FRM
overestimates diameters and underestimates albedos. At spin axis
latitudes close to the extremes (-90\degr\ and +90\degr), the largest
fraction of the insolation falls near the equator, leading to a
surface temperature distribution very similar to that of the FRM,
penalizing the NEATM performance. The wavy pattern in the case of the
spin axis longitude is caused by our selection effects for the orbital
geometries: all geometries used in this simulation have the same epoch
and solar elongations between 60\degr\ and 180\degr\ (see Section
\ref{lbl:space}). For those synthetic objects with spin axis latitudes
close to zero, the spin axis longitude decides whether insolation is
focused on the poles or the equator. Hence, NEATM-preferred and
FRM-preferred surface temperature distributions are spaced by 90\degr\
in spin axis longitude, causing a wave-like pattern in the fractional
errors of this parameter.

Despite the complexity caused by the different spin axis
configurations, the NEATM is more likely to provide accurate diameters
and albedos for a random spin axis orientation of a random NEA.

\subsection{Observing Geometry}
\label{lbl:geometry}

\begin{figure}
\epsscale{1}
\plotone{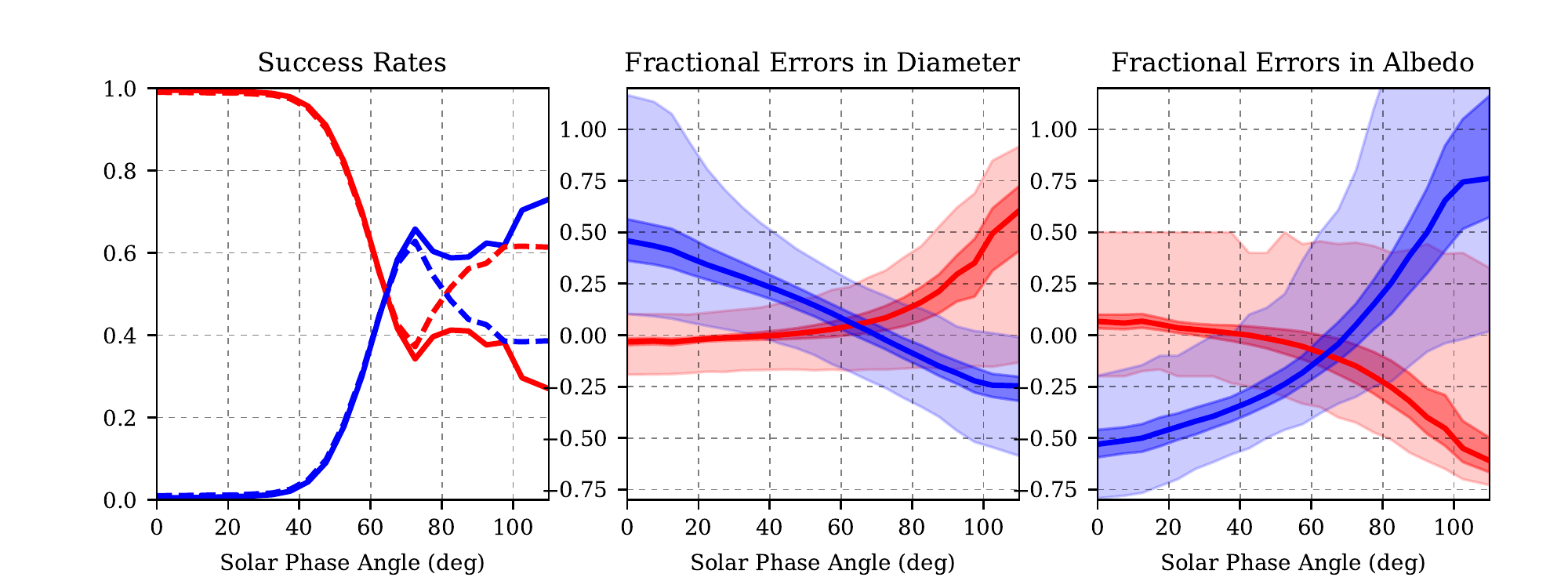}
\caption{Success rates and fractional errors based on solar phase
  angle. Refer to Figure \ref{fig:physicalproperties} for
  definitions. There is a strong dependence of the success rates and
  fractional errors of the NEATM and the FRM as a function of solar
  phase angle. FRM diameters are generally closer to the real
  diameters than NEATM diameters for phase angles greater
  65\degr\label{fig:geometry}}
\end{figure}

We investigate the impact of the synthetic objects' observing geometry
on the thermal modeling results. Since heliocentric distance, distance
from the observer, and solar phase angle are closely related, we focus
on the latter, which shows the effects most clearly (Figure
\ref{fig:geometry}). At solar phase angles up to 65\degr, the NEATM
has a success rate close to unity and its fractional errors in
diameter and albedo are close to zero. For phase angles greater than
65\degr, the FRM diameter success rate is greater than that of the
NEATM. The situation is more complicated for the albedo success rates
of the NEATM and the FRM. This behavior is caused by the fact that the
absolute diameter fractional errors of the NEATM grow faster than
those of the FRM; FRM diameters are closer to the real diameters for
solar phase angles greater 65\degr. On the other hand, the absolute
albedo fractional errors of the NEATM grow slower than those of the
FRM; for solar phase angles $65\degr \leq \alpha \leq 80\degr$ the FRM
derives on average slightly more accurate albedos than the NEATM, but
for $\alpha > 80\degr$, the NEATM performs slightly better again,
causing the fluctuations in the albedo success rates.

The overall trend in the diameter and albedo fractional errors is a
consequence of the oversimplified assumptions going into the thermal
models. NEATM's failing at phase angles greater than 65\degr\ is a
result of its lack of night side emission. As the solar phase angle
increases, a larger fraction of the asteroid's night side, which is
assumed to not contribute to its thermal emission, is
observed. Observed flux densities are greater than modeled flux
densities, leading to an overestimation of the object's diameter.  The
fact that the FRM overestimates the diameter for solar phase angles
less than 65\degr\ and underestimates it for greater phase angles
stems from its limited capabilities to adapt more complex surface
temperature distributions. At low solar phase angles, the observer
sees the subsolar point, which is usually hotter than any temperature
the FRM assumes, leading to an overestimation of the diameter. For
high solar phase angles, the subsolar point is below the horizon and a
large fraction of the night side is observed, underestimating the
diameter. For both models, albedos are overestimated if diameters are
underestimated and vice-versa.


\subsection{Statistical Correction Functions}
\label{lbl:corrections}

\begin{figure}
\epsscale{1}
\plotone{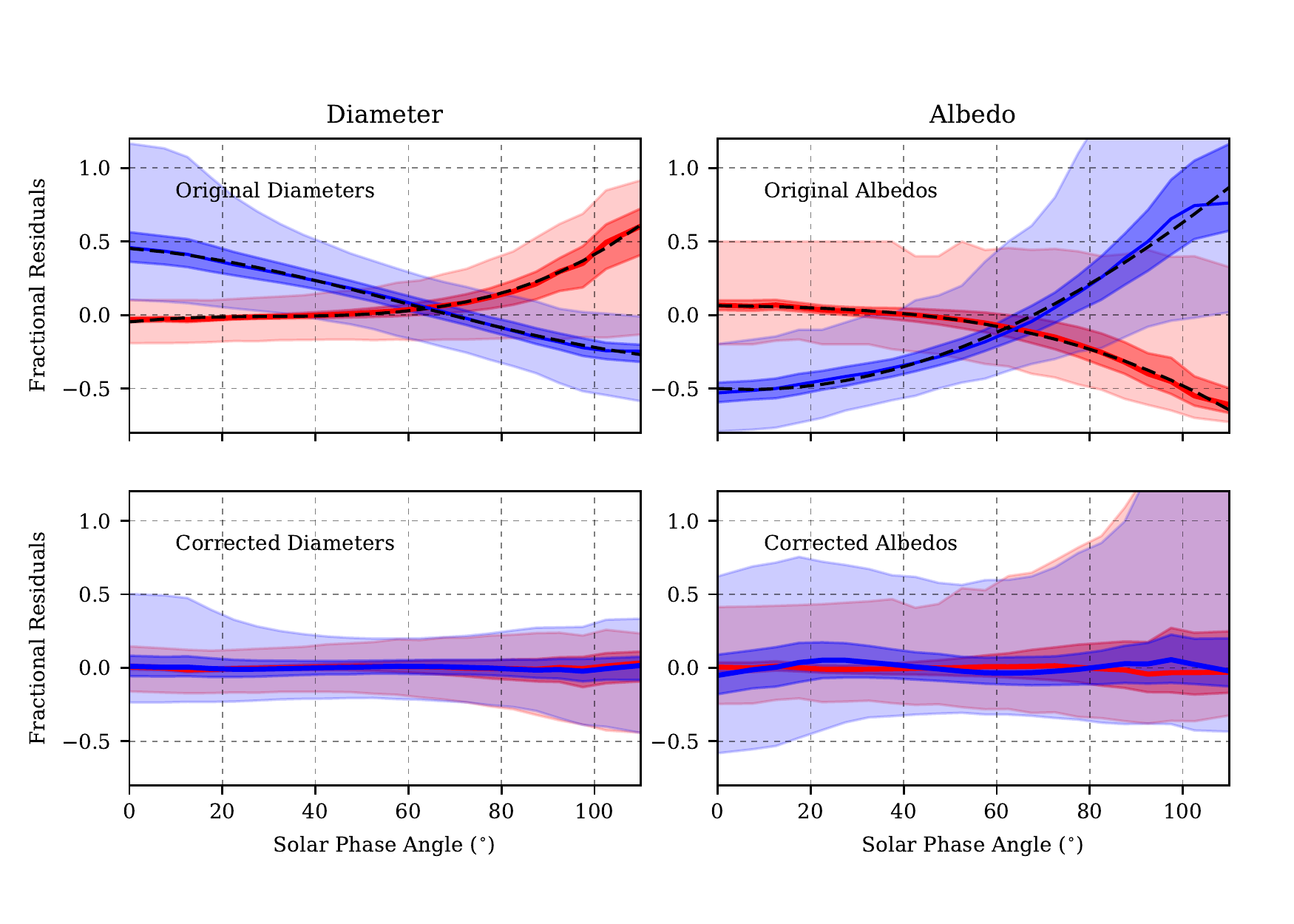}
\caption{{\bf Top}: NEATM and FRM fractional errors in diameter
  (left) and geometric albedo (right). The same data are shown as in
  Figure \ref{fig:geometry}. The dashed lines are statistical offsets
  derived through third order polynomial fits to the median fractional
  errors as a function of solar phase angle. {\bf Bottom}: The same
  distributions as shown in the top panel after the statistical offsets
  have been subtracted from the derived diameters/albedos on a
  per-object basis. The median fractional errors of the corrected
  diameters and albedo are close to zero over the entire phase angle
  range.\label{fig:corrections}}
\end{figure}

We take advantage of the systematic effect of the solar phase angle on
the median fractional errors to establish statistical correction
functions for diameter and albedo estimates derived with the NEATM and
the FRM for individual objects. The correction functions are
third-order polynomials that are fitted to the median fractional
errors shown in Figure \ref{fig:corrections} (top panel) using a
$\chi^2$ minimization technique; the best-fit polynomial coefficients
are listed in Table \ref{tab:corrections} and the fits are plotted in
the top panel of Figure \ref{fig:corrections}. The correction
functions are of the form $f(\alpha)= \sum_{i=0}^{3} c_i \alpha^i$
with solar phase angle $\alpha$ measured in degrees. In order to
derive corrected diameters ($d_{\mathrm{corr}}$) from NEATM diameters
($d_{\mathrm{NEATM}}$), use the following relation:

\begin{equation}
  d_{\mathrm{corr}}= \frac{d_{\mathrm{NEATM}}}{1+f(\alpha)},
\end{equation}

where $f(\alpha)$ is the corresponding NEATM diameter function and
$\alpha$ the solar phase angle in degrees. FRM and albedo corrections
are performed correspondingly. The corrections can be applied to
  individual objects. The limitations of this approach are discussed
  in Section \ref{lbl:discussion}.

\begin{deluxetable}{rcccc}
\tablecaption{Polynomial Coefficients of Statistical Diameter
   and Geometric Albedo Correction Functions. \label{tab:corrections}}
\tablewidth{0pt} 
\tablehead{
  \colhead{Model} &
  \colhead{$c_0 (10^{-2})$} &
  \colhead{$c_1 (10^{-4})$} &
  \colhead{$c_2 (10^{-6})$} &
  \colhead{$c_3 (10^{-7})$}}
\startdata
NEATM diameter & -4.7 & 33.4 & -105.6 & 11.8 \\
FRM diameter  & 45.3 & -24.7 &  -95.9 & 5.3 \\
\hline
NEATM albedo & 6.4 & -5.2 & -2.1 & -4.7 \\
FRM albedo & -50.0 & -19.8 & 148.7 &  -1.6 \\
\enddata
\end{deluxetable}

The bottom panel of Figure \ref{fig:corrections} shows the median
fractional errors for the NEATM and the FRM after the statistical
corrections have been applied to each individual synthetic object on a
per-object basis. Over the entire phase angle range, the median
corrected diameters and albedos agree with the actual values
(fractional error equals zero) within their 1$\sigma$ envelopes.

\section{Discussion}
\label{lbl:discussion}

In Section \ref{lbl:results} we presented the effects of individual
physical and orbital properties on thermal modeling results. The
largest effect is caused by the target's solar phase angle. While this
effect was already known before \citep[see, e.g.,][]{Delbo2004,
  Wright2007}, we quantified the magnitude of this effect in terms of
the success rate and fractional errors and provided a correction
function for both models. We also found that thermal modeling results
of the NEATM and the FRM are barely affected by other parameters like
thermal inertia, rotational period, geometric albedo, surface
roughness, and pole orientation.

The parameter space explored in this work has been chosen to mimic the
properties of NEAs to the best of our knowledge. While useful
estimates are available for the distributions of geometric albedos and
spin axis latitudes, no un-biased distributions are currently
available for rotational period, thermal inertia, surface roughness,
and spin axis longitude. For the other parameters, it is our intention
to sample them in a wide range resembling the observed range in a
uniform way. We are aware that the random combination of parameters
might lead to some unrealistic configurations (e.g., a large
low-albedo asteroid with a rotational period of less than a minute);
we deliberately refrain from rejecting such cases in order to not
distort the results of this study.

We investigate the impact of our choice of parameter ranges on the
results of this work by splitting the uniform distributions in two
equally sized parts at their respective median values (e.g., in the
case of rotational period: short periods with $P\leq50$~hr and long
periods with $P>50$~hr) and repeat the analysis discussed in Section
\ref{lbl:results} on each of these sub-samples. While these sub-sample
distributions do not describe reality better than the original
distribution, this approach allows us to explore the impact of a
entirely different distribution on our results. We find that the
results for each sub-sample agrees well with those shown in Figures
\ref{fig:physicalproperties} to \ref{fig:geometry}. Differences in
fractional errors are in most cases within the 1$\sigma$ envelopes
-- in all cases they are well-within the 3$\sigma$ envelopes. In
general, FRM diameters and albedos seem to be more affected by the
sub-sample parameter distributions, which is likely due to its
inability to adjust its surface temperature distribution in the way
the NEATM does by varying the beaming parameter $\eta$. In order to
quantify the differences, we investigate the median fractional errors and the
root-mean-square residuals between the statistical correction
functions derived in Section \ref{lbl:corrections} and the individual
sub-samples over the full range in solar phase angle (0 to
110\degr). We find median errors to be less than 5\% and
root-mean-square residuals to be less than 10\% for all sub-samples
over the full range in phase angle. Again, FRM errors tend to be
higher for the aforementioned reasons. While these errors are not
negligible, they are well within the typical uncertainties of thermal
models \citep[e.g.,][20\% in diameter, 50\% in albedo]{Harris2011},
proving the usefulness of our approach.

Our approach in this analysis is simplistic in the sense that we only
investigate effects of individual parameters, keeping the underlying
distributions of all other parameters intact. However, specific
combinations of the aforementioned parameters are able to
significantly distort thermal modeling results. This is reflected by
success rates that deviate from unity and the wide 3$\sigma$
confidence intervals of the fractional errors plotted in Figures
\ref{fig:physicalproperties} to \ref{fig:geometry}. For instance,
Figure \ref{fig:physicalproperties} shows the effects of rotational
period and thermal inertia on thermal modeling results separately,
both of which only have a minor impact. However, it is likely that for
a combination of a short rotational period and a high thermal inertia
(and some other parameters) the FRM can perform better than the
NEATM. Cases like these are captured by the fractional error envelopes
that we derive.  We deliberately do not investigate the effect of
combinations of physical and orbital parameters, since this study is
aimed at the thermal modeling of generic NEAs for which very little
information is available. In cases in which more information is
available (e.g., rotational periods, spin axis orientations, etc.) on
the target body, the use of a TPM might be more appropriate and
provide more accurate results.

The statistical correction functions derived in Section
\ref{lbl:corrections} have been derived from large samples of
synthetic data and might not have the effect of improving the diameter
estimates for individual asteroids. The corrections are meant to
provide an estimate of how much the different thermal models
overestimate or underestimate diameters and albedos on average. They provide a
means to correct for thermal model limitations on a statistical
basis. Some of these limitations are the FRM's inability to modulate
its surface temperature distribution and the NEATM's assumption of a
lack of night side thermal emission. A different approach to the
latter problem has been taken by \citet{Wolters2009} by introducing
the ``Night Emission Simulated Thermal Model,'' which assumes a fixed
surface thermal inertia that allows for an approximate determination
of the thermal emission from the night side. While the NESTM has been
found to provide better diameter accuracy than the NEATM for solar
phase angles greater than 45\degr, it relies on a number of
assumptions (fixed rotational period and thermal inertia) that may
reduce its applicability to a wide range of asteroid observations. The
NESTM and our statistical correction function are different approaches
to the same problem, both of which may be limited in their effect on
individual asteroid observations.

\subsection{Implications for Existing Diameter and Albedos
  Measurements}

Diameters and albedos for about 2,500 NEAs have been derived from
thermal-infrared observations. The majority of these observations have
been performed during the previous decade, namely by the {\it Spitzer
  Space Telescope} \citep{Trilling2010, Trilling2016} and the {\it
  Wide-field Infrared Survey Explorer} \citep[e.g.,][]{Mainzer2011b}, and diameters and albedos have been derived using the
NEATM. Due to spacecraft design constraints, most observations were performed
near quadrature, leading to similar constraints on the solar phase
angle for NEA observations from both observatories. Both the Spitzer
programs (ExploreNEOs, NEOSurvey, and NEOLegacy; see the SpitzerNEOs
database \url{http://nearearthobjects.nau.edu/spitzerneos.html}) and
the WISE observations \citep{Mainzer2011b} had a significant fraction
of their targets observed at solar phase angles greater than 65\degr\,
but generally less than 90\degr. According to Figure
\ref{fig:geometry}, this implies that for these observations,
diameters are likely to be overestimated by up to 25\% and albedos
underestimated by up to 40\% (relative). These effects are of the same
order of magnitude as the NEATM accuracies derived by
\citet{Harris2011} and hence represent a systematic effect that is not
negligible.

We estimate the magnitude of this effect on the ensemble of
radiometric NEA diameters and albedos available to date by querying
the SpitzerNEOs database (October 2017) and find that about 25\% of
all Spitzer-observed NEAs were observed at solar phase angles greater
than 65\degr; ${\sim}5$\% were observed at $\alpha > 80$\degr\ and
${<}0.5$\% were observed at $\alpha > 90$\degr. A similar analysis for
WISE-observed NEAs is not possible as observing geometries for these
targets have not been published. However, according to Figure 7 in
\citet{Mainzer2011b}, the ratios seem to be similar. Hence, we
conclude that a small fraction (${<}5$\%, the fraction of
Spitzer-observed NEAs with $\alpha > 80$\degr) of existing NEA
diameter and albedo estimates are likely to be affected by systematic
effects that are of the same order of magnitude as the expected NEATM
accuracy \citep{Harris2011}. We further strongly encourage researchers
to include information on the observing geometry with future
publications using thermal models.

\subsection{Limitations}

Our study is limited by the fact that all synthetic asteroids are
assumed to be spherical and homogeneous in their
properties. Furthermore, we assume no noise model in the derivation of
the synthetic asteroid flux densities (flux density uncertainties
following Poisson noise have been derived for the $\chi^2$
minimization in the thermal models (Section \ref{lbl:modeling}); note
that the generated fluxes have not been modified based on these
uncertainties). While these simplifications allow us to better probe
the effects we wanted to study, it is a gross simplification of real
asteroid observations. Hence, we deliberately do not suggest that the
fractional error confidence intervals shown in Figures
\ref{fig:physicalproperties} to \ref{fig:corrections} provide a
measure of the accuracies that the NEATM and the FRM can provide for
real asteroid observations. The result that the NEATM fractional
error 1$\sigma$ confidence intervals are of the order of 10\% in
most cases provides merely a lower limit to the diameter accuracy that
can be reached by the NEATM under idealized conditions that minimize
uncertainties in the measurement process and the object's
lightcurve. In light of this lower limit, a 20\% uncertainty on
diameters derived with the NEATM, as derived by \citet{Harris2011},
seems to be realistic. A similar argument for the albedo uncertainties
is not useful, since albedo uncertainties are mainly driven by the
accuracy of the target's absolute magnitude in the optical
\citep{Harris2002} and lightcurve effects.

The results of this study are also limited by the applicability of the
TPM, which does not consider lateral heat conduction and is only valid
in cases in which the diameter of the body is significantly larger
than the thermal skin depth, which is usually of the order of a few
centimeters \citep{Spencer1989, Mueller2007}. Hence, the results of
this work are valid for meter-sized and larger asteroids. However,
future observations of decimeter-sized bodies in close encounters with
Earth or so-called minimoons might require additional modeling for a
full interpretation of the results that can be obtained from
thermal-infrared observations.

\subsection{Discussion of Individual Cases}

\begin{figure}
\epsscale{0.8}
\plotone{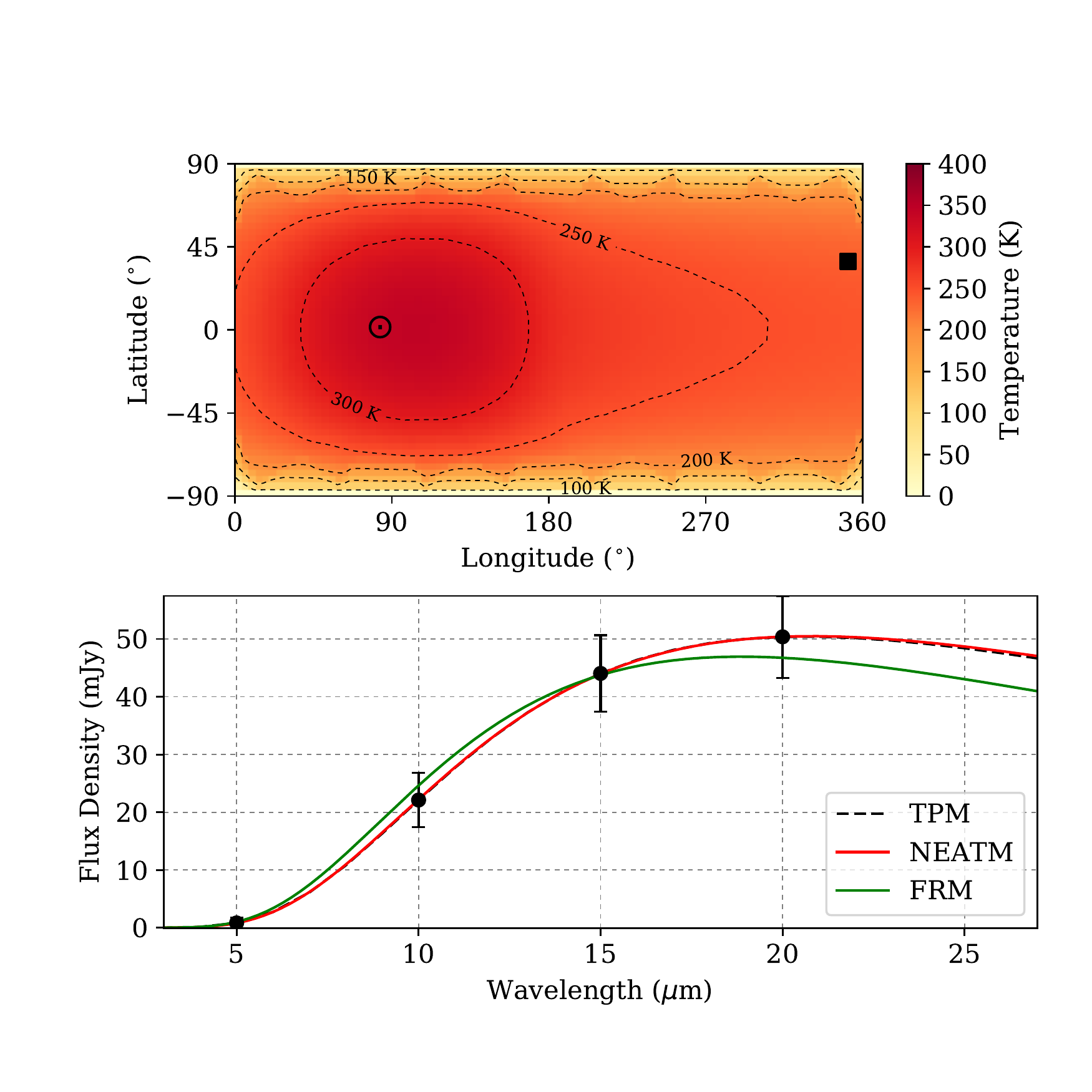}
\caption{Surface temperature distribution of a random synthetic NEA
  observed at high solar phase angle (compare to Figure
  \ref{fig:neatm_frm_temps}.) The subsolar point is indicated by a Sun
  symbol; the sub-observer point is indicated by a black square. For
  this example (rotational period of 0.55~hr, thermal inertia of
  200~SI units), the FRM provides a much better diameter estimate
  (0.37~km instead of the real diameter 0.45~km; geometric albedo:
  0.81 instead of the real geometric albedo 0.55) than the NEATM
  (0.78~km, geometric albedo: 0.18), although the NEATM fits the
  object's SED much better than the FRM (bottom panel). This
  discrepancy is a result of the fact that the NEATM assumes zero
  thermal emission from the night-side of the object, which is clearly
  not the case as shown in the top panel, leading to an overestimation
  of the object's diameter. \label{fig:example_high_alpha}}
\end{figure}

\begin{figure}
\epsscale{1}
\plotone{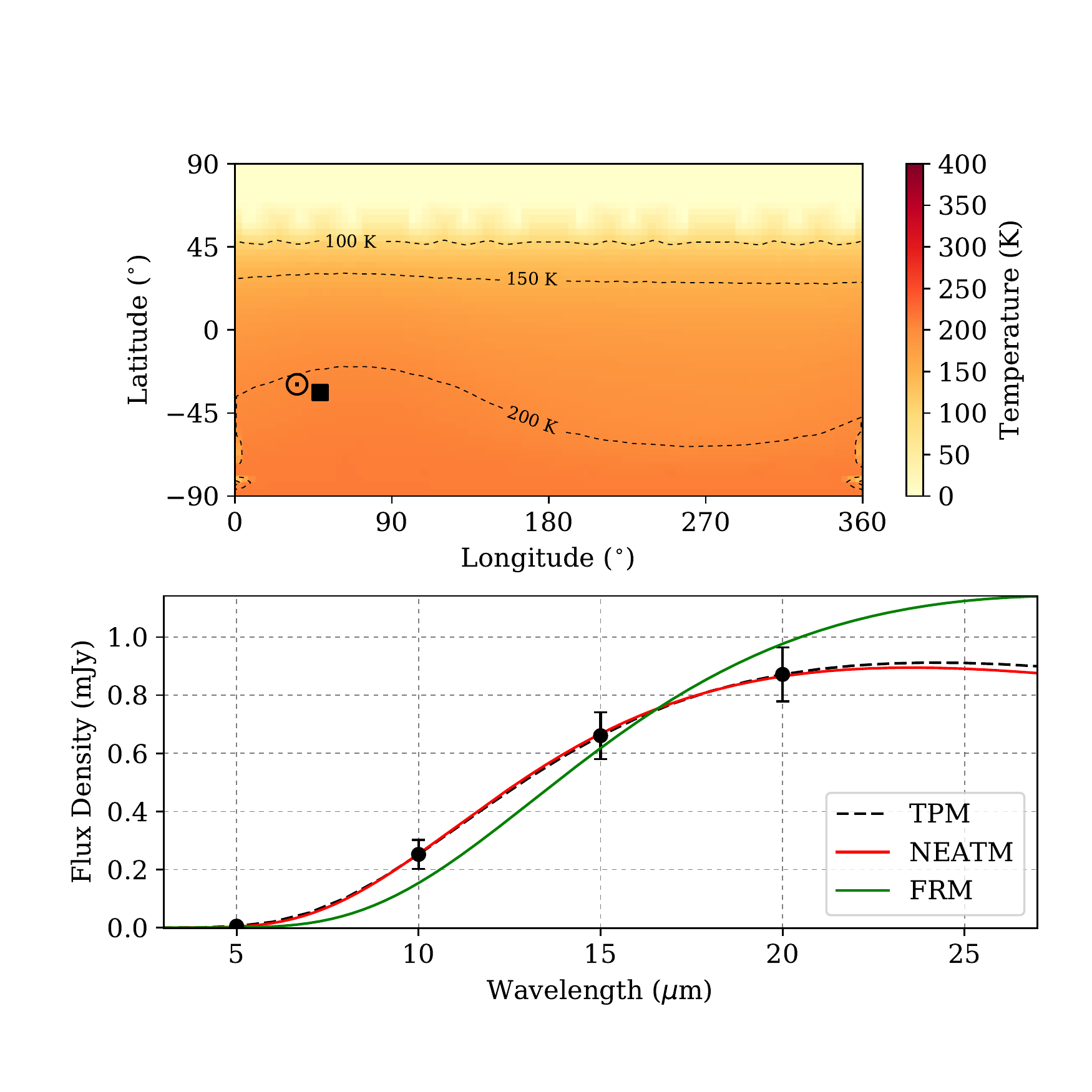}
\caption{Surface temperature distribution of a random synthetic NEA
  observed at low solar phase angle (compare to Figure
  \ref{fig:neatm_frm_temps}.) The subsolar point is indicated by a Sun
  symbol; the sub-observer point is indicated by a black square. For
  this example (rotational period of 0.28~hr, thermal inertia of
  458~SI units), the NEATM provides much better results (diameter:
  0.6~km, real diameter: 0.63~km; geometric albedo: 0.3, real
  geometric albedo: 0.28) than the FRM (diameter: 0.86~km, geometric
  albedo: 0.15), and it also provides the better fit to the object's
  SED. Note how the NEATM provides a better estimate of the object's
  diameter, even though the NEA's surface temperature distribution is
  nearly iso-latitudinal. \label{fig:example_low_alpha}}
\end{figure}

Figures \ref{fig:example_high_alpha} and \ref{fig:example_low_alpha}
illustrate two cases of synthetic asteroids observed at large and
small solar phase angles, respectively, and explain the superiority of
the FRM in the former and the NEATM in the latter case.

Figure \ref{fig:example_high_alpha} demonstrates why the FRM provides
more accurate model diameters than the NEATM at high phase angles. The
NEATM's assumption of zero night-side emission plays an increasing
role with increasing solar phase angle. At large phase angles, the
NEATM assumes that a large fraction of the observed surface area is on
the object's night side and hence cold -- for solar phase angles
greater 90\degr, the subsolar point is not visible to the
observer. For the synthetic body depicted in Figure
\ref{fig:example_high_alpha}, the night-side surface temperature is
much higher than assumed by the NEATM. Hence, the
NEATM tries to compensate for the lack of night-side emission by
increasing the model diameter, yielding an overestimation of the
object's diameter. In this case, the FRM provides a better
approximation of the surface temperature distribution and the object's
diameter. 

Interestingly, the NEATM provides a much better fit to the
object's SED than the FRM. This is the result of the NEATM's ability
to modulate the subsolar temperature to fit the observations. However,
in this case, a good fit does not equal a good approximation of the
object's diameter, since the assumptions on which the model is built
describe this configuration badly. Hence, the simpler FRM provides a
better approximation of this situation. The ideal solution to this
problem would be to utilize a TPM to solve for the target's diameter
and albedo. However, this approach requires additional information on
the target's spin axis orientation, rotation period, and other
properties. Since this wealth of information is usually not available,
we suggest the comparison of NEATM and FRM results, as these will most
likely bracket the real diameter and albedo.

In Figure \ref{fig:example_low_alpha}, we present the surface
temperature distribution of a random synthetic asteroid observed at
low solar phase angles. Although the object's surface temperature
distribution is nearly iso-latitudinal, the NEATM better describes the
object's diameter, which is at least in part due to the fact that the
subsolar point (the surface's hottest point) lies within the observed
hemisphere, as is assumed for the NEATM for small phase angles. Hence,
low phase angle observations are usually better described with the
NEATM, even though the temperature distribution is somewhat close to
that of the FRM. This observation can be attributed to the fact that
real asteroids have non-zero thermal inertias, finite rotational
periods, and spin axis orientations that are different from the
equator-on assumption on which the NEATM is built, leading to a
maximum surface temperature that is lower than predicted by the
subsolar temperature.

\section{Conclusions}

We investigate the effects of physical and orbital properties on the
accuracy of diameters and albedos derived with the NEATM and the
FRM. This analysis is done by simulating a synthetic but realistic
population of NEAs and comparing the diameters and albedos derived
with the NEATM and the FRM with the individual input parameters. We
find the NEATM to provide more accurate diameters and albedos than the
FRM in most cases, with a few exceptions. The modeling results are
barely affected by the physical properties of the objects (rotational
period, thermal inertia, spin axis orientation, surface roughness, and
geometric albedo), but we find a large impact of the solar phase angle
on the modeling results. We conclude that the NEATM provides
statistically more robust diameter estimates for NEAs observed at
solar phase angles less than ${\sim}$65\degr, while the FRM provides
more robust diameters estimates for solar phase angles greater than
${\sim}$65\degr. This finding implies that a small percentage
(${<}5$\%) of all available NEA diameters (albedos) to date are
overestimated by up to 25\% (underestimated by up to 40\%). We provide
statistical correction functions for diameters and albedos derived
with the NEATM and the FRM as a function of solar phase angle. The
primary conclusion of this work is that errors induced by the
application of the NEATM and the FRM need to be carefully modeled
based on the observations in order to characterize the resulting
uncertainties, especially at large phase angles.

\acknowledgments

The authors would like to thank an anonymous referee for providing
useful suggestions that improved this manuscript.  Some of this
research was supported by NASA's Solar System Evolution Research
Virtual Institute (SSERVI) as part of the Institute for the Science of
Exploration Targets (ISET) at the Southwest Research Institute (NASA
grant no.\ NNA14AB03A). MM was supported in part by NASA grant
No. NNX12AG07G from the Near Earth Object Observations program. RJ was
supported in part by internal funding from the University of Hawai`i's
Institute for Astronomy and the Office of the Vice Chancellor for
Research. Computational analyses were run on Northern Arizona
University's Monsoon computing cluster, funded by Arizona's Technology
and Research Initiative Fund.

\end{document}